\documentclass[reprint, amsmath, amssymb, floatfix, aps, superscriptaddress, nofootinbib, nobibnotes, onecolumn]{revtex4-1}
\usepackage{amsfonts}
\usepackage{mathrsfs}
\usepackage{natbib}
\usepackage{graphicx}
\usepackage{hyperref}
\usepackage{bm}
\usepackage{subfig} 
\usepackage{amsmath}
\DeclareMathAlphabet{\mathscrbf}{OMS}{mdugm}{b}{n}
\usepackage{geometry}
\usepackage{indentfirst}
\usepackage{color}
\usepackage{acro}
\usepackage{svg}
\usepackage{tablefootnote}

\DeclareAcronym{GW}{
  short = GW ,
  long = gravitational wave ,
  short-plural = s 
}
\DeclareAcronym{LIGO}{
  short = LIGO ,
  long = Laser Interferometer Gravitational-wave Observatory ,
  short-plural = 
}
\DeclareAcronym{LISA}{
  short = LISA ,
  long = Laser Interferometer Space Antenna ,
  short-plural =  
}
\DeclareAcronym{SKA}{
  short = SKA ,
  long = Square Kilometre Array ,
  short-plural =  
}  
  
\DeclareAcronym{SNR}{
	short = SNR ,
	long = signal-to-noise ratio ,
	short-plural = 
}

\DeclareAcronym{PTA}{
	short = PTA ,
	long = pulsar timing array ,
	short-plural = 
}

\DeclareAcronym{PDF}{
	short = PDF ,
	long = probability density function ,
	short-plural = s
}

\DeclareAcronym{FLRW}{
  short = FLRW ,
  long = Friedmann-Lemaitre-Robertson-Walker ,
  short-plural =  
}

\DeclareAcronym{SIGW}{
	short = SIGW ,
	long = scalar induced gravitational wave ,
	short-plural =  s
}

\DeclareAcronym{PBH}{
	short = PBH ,
	long = primordial black holes ,
	short-plural =  s
}

\DeclareAcronym{CMB}{
	short = CMB ,
	long = cosmic microwave background ,
	short-plural =  s
}
\DeclareAcronym{DM}{
	short = DM ,
	long = dark matter ,
	short-plural =  s
}

\DeclareAcronym{LSS}{
	short = LSS ,
	long = large scale structure ,
	short-plural =  s
}

\DeclareAcronym{RD}{
	short = RD ,
	long = radiation-dominated ,
	short-plural =  
}

\begin{document}

\title{Primordial black holes and scalar induced density perturbations: the effects of probability density functions}

\author{Jing-Zhi Zhou} 
\affiliation{Center for Joint Quantum Studies and Department of Physics,
School of Science, Tianjin University, Tianjin 300350, China}

\author{Yu-Ting Kuang} 
\email{kuangyt@ihep.ac.cn}
\affiliation{Institute of High Energy Physics, Chinese Academy of Sciences, Beijing 100049, China}
\affiliation{University of Chinese Academy of Sciences, Beijing 100049, China}

\author{Zhe Chang} 
\affiliation{Institute of High Energy Physics, Chinese Academy of Sciences, Beijing 100049, China}
\affiliation{University of Chinese Academy of Sciences, Beijing 100049, China}

\author{Xukun Zhang}   
\affiliation{College of Physics and Information Engineering, Shanxi Normal University, Taiyuan 030031, China}

\author{Qing-Hua Zhu}   
\affiliation{Department of Physics and Chongqing Key Laboratory for Strongly Coupled Physics, Chongqing University, Chongqing 401331, China}


\begin{abstract}
We investigate the second order energy density perturbation $\delta^{(2)}$ induced by small-scale Gaussian and local-type non-Gaussian primordial curvature perturbations. The relative abundance of \acp{PBH} is calculated in terms of the \acp{PDF} of total energy density perturbation $\delta_r=\delta^{(1)}+\frac{1}{2}\delta^{(2)}$. The effects of second order density perturbation greatly reduce the upper bounds of small-scale power spectra of primordial curvature perturbations by one to two orders of magnitude. For log-normal primordial power spectrum, its amplitude $A_{\zeta}$ is constrained to be about $A_{\zeta}\sim 3\times10^{-3}$. And for local-type non-Gaussianity with $f_{\mathrm{NL}}=10$, the upper bound of $A_{\zeta}$ is about $2.5\times10^{-4}$.
\end{abstract}

\maketitle

\section{Introduction}\label{sec:intro}
In inflation theory, valuable information about the early Universe is encoded in the cosmological perturbations which are originated from the quantum fluctuations during inflation. The power spectrum of primordial curvature perturbations $\mathcal{P}_{\zeta}\left(k \right)$ is one of the most important predictions in inflation theory. The amplitude of the primordial power spectrum at different scales can be constrained by current cosmological observations. On large scales ($\gtrsim$1 Mpc), according to the current observations of the \ac{CMB} and \ac{LSS} \cite{Planck:2018vyg,Abdalla:2022yfr}, the amplitude of the power spectrum of primordial curvature perturbations is constrained to be about $2\times 10^{-9}$. However, on small scales ($\lesssim$1 Mpc), the constraints of primordial curvature perturbations are significantly weaker than those on large scales \cite{Bringmann:2011ut}. Constraining the primordial power spectrum on small scales is an important issue in the study of early Universe.

There are two approaches to constrain or probe the primordial power spectrum on small scales: primordial black hole abundance and scalar induced gravitational waves. More precisely, the large amplitudes of small-scale primordial spectrum have been attracting a lot of attention in the past few years on account of their rich and profound phenomenology, such as \ac{PBH}  \cite{Sasaki:2018dmp,Carr:1975qj,Carr:1993aq,Carr:2016drx,Carr:2020gox,Carr:2020xqk,Garcia-Bellido:2017mdw,Ivanov:1994pa,DeLuca:2020agl,DeLuca:2020qqa,Vaskonen:2020lbd,Braglia:2020eai,Bartolo:2018evs,Byrnes:2018clq,Ballesteros:2017fsr,Ezquiaga:2017fvi,Inomata:2016rbd,Kawasaki:2016pql,Young:2013oia,Gow:2020bzo,Wang:2021djr,Khlopov:2008qy,Belotsky:2018wph,Wang:2016ana,Wang:2019kaf} which is a candidate of \ac{DM}. The abundance of \acp{PBH} $f_{\mathrm{pbh}}$ can be calculated in terms of a given primordial power spectrum of curvature perturbation $\mathcal{P}_{\zeta}(k)$. The abundance of \acp{PBH}, $f_{\mathrm{pbh}}(m_{\mathrm{pbh}})$, has been constrained on different masses of \acp{PBH} $m_{\mathrm{pbh}}$ \cite{Carr:2009jm,Carr:2020gox}, then the corresponding small-scale primordial power spectrum can also be constrained by various observations of \acp{PBH}. Furthermore, the primordial perturbations will inevitably generate higher order perturbations, such as \acp{SIGW}  \cite{Domenech:2021ztg,Mollerach:2003nq,Ananda:2006af,Baumann:2007zm,Kohri:2018awv,Zhou:2021vcw,Chang:2022nzu,Saito:2008jc,Wang:2019kaf,Inomata:2018epa,Byrnes:2018txb,Garcia-Bellido:2017aan,Nakama:2016gzw,Bugaev:2010bb,Saito:2009jt,Bugaev:2009zh,Inomata:2020lmk,Ballesteros:2020qam,Lin:2020goi,Chen:2019xse,Cai:2019elf,Cai:2019jah,Ando:2018qdb,Di:2017ndc,Chang:2022dhh,Chang:2022vlv,Orlofsky:2016vbd,Gao:2021vxb,Nakama:2015nea,Nakama:2016enz,Changa:2022trj,Zhao:2022kvz}. This strong correlation between primordial curvature perturbation and \acp{SIGW} signals might be a promising approach to detecting small-scale primordial power spectrum in the \ac{GW} experiments, such as \ac{LISA} and \ac{PTA} \cite{NANOGrav:2023hvm,Wang:2023ost,Bi:2023tib,Yuan:2019udt,Thrane:2013oya,Robson:2018ifk,Kuroda:2015owv,Siemens:2013zla,Zhou:2024doz,Zhao:2022kvz,Chang:2023vjk,Chang:2023aba,Ellis:2023oxs,Franciolini:2023pbf,Domenech:2024rks}.

In this paper, we investigate the contributions of second order induced energy density perturbations $\delta^{(2)}=\rho^{(2)}/\rho^{(0)}$ to the \acp{PBH} formation. More precisely, the primordial curvature perturbation that enters the Hubble radius during the radiation-dominated (RD) era will lead to the generation of higher order scalar and energy density perturbations. Then, the relative abundance of \acp{PBH} can be studied in terms of the \acp{PDF} of the total energy density perturbations:$P\left(\delta^{(1)}+\frac{1}{2}\delta^{(2)}\right)$, where $\delta^{(1)}$ is the contributions of the primordial curvature perturbation which have been studied for many years \cite{Carr:2020xqk,Sato-Polito:2019hws}. Here, $\delta^{(2)}$ is the new contribution of second order scalar induced energy density perturbation $\delta^{(2)}$ which was neglected in previous studies of primordial black holes\footnote{ The statistics of induced higher order energy density perturbation $\delta^{(2)}$ are highly non-Gaussian, since they are generated by first order scalar perturbations.}. Since the abundance of \acp{PBH} $f_{\mathrm{pbh}}(m_{\mathrm{pbh}})$ has been constrained on different $m_{\mathrm{pbh}}$, the amplitude of small-scale primordial power spectra can be constrained by the abundance of \acp{PBH}.

This paper is organized as follows. In Sec.~\ref{sec:2.0}, we calculate the second order scalar induced density perturbation in comoving gauge. In Sec.~\ref{sec:3.0}, we study the impact of second-order scalar-induced density perturbations on the PDF function and  discuss the constraints on the primordial black hole abundance imposed by the small-scale primordial power spectrum. Finally, we summarize our results and give some discussions in Sec.~\ref{sec:4.0}.

\section{Second order induced energy density perturbation}\label{sec:2.0}
The perturbed metric in the \ac{FLRW} spacetime with comoving gauge takes the form
\begin{eqnarray}
	\mathrm{d}s^{2}&&=a^{2}\left(-\left(1+2 \phi^{(1)}+ \phi^{(2)}\right) \mathrm{d} \eta^{2}+\left(2 \partial_iB^{(1)}+\partial_iB^{(2)} \right) \mathrm{d} \eta \mathrm{d} x^{i} \right. \nonumber\\
	&&\left.+\left(1-2 \psi^{(1)}- \psi^{(2)} \delta_{i j}\right)\mathrm{d} x^{i} \mathrm{d} x^{j}\right)\ ,
\end{eqnarray}
where $\phi^{(n)}$, $\psi^{(n)}$, and $B^{(n)}$$(n=1,2)$ are the $n$th-order scalar perturbations. We have set  the $n$-th order perturbation of the scalar part of the velocity, $\delta u^{(n)}=0$, and the scalar metric perturbation, $E^{(n)}=0$, in comoving gauge. In the radiationdominated (RD) era, the first order scalar perturbation is given by
$$
\begin{aligned}
	\psi(\eta, \mathbf{k}) =\frac{2}{3} \zeta_{\mathbf{k}} T_\phi(k \eta) \ , \ \phi(\eta, \mathbf{k})=\frac{2}{3} \zeta_{\mathbf{k}} T_\psi(k \eta)\ , \	B(\eta, \mathbf{k}) =\frac{2}{3 k} \zeta_{\mathbf{k}} T_B(k \eta)\ ,
\end{aligned}
$$
where $\zeta_{\mathbf{k}}$ is the primordial curvature perturbations. The transfer functions $T_\phi(k \eta), T_\psi(k \eta)$, and $T_B(k \eta)$ in the RD era are \cite{Lu:2020diy}
$$
\begin{aligned}
	T_\psi(x)= & \frac{3}{2} \frac{\sin (x / \sqrt{3})}{x / \sqrt{3}} \ , \\
	T_\phi(x)= & \frac{3}{2}\left(\frac{\sin (x / \sqrt{3})}{x / \sqrt{3}}-\cos (x / \sqrt{3})\right)\ , \\
	T_B(x)=&\frac{3}{2 x^2}\left(6 x\cos (x/ \sqrt{3})+\sqrt{3}\left(x^2-6\right) \sin (x / \sqrt{3})\right) \ ,
\end{aligned}
$$
where we have defined $x \equiv k \eta$. The higher order cosmological perturbations can be studied in terms of the \texttt{xPand} package \cite{Pitrou:2013hga,Tomikawa:2019tvi}. The equations of motion of second order scalar perturbations are
\begin{eqnarray}\label{eq:1}
	 2\mathcal{H}\phi^{(2)'}+6\mathcal{H}\psi^{(2)'}+2\psi^{(2)''}&&+\frac{8}{3}\mathcal{H}\Delta B^{(2)}+\Delta B^{(2)'}  \nonumber\\
	&&+\Delta \phi^{(2)}-\frac{5}{3}\Delta \psi^{(2)}=- \mathcal{T}^{ij} S^{(2)}_{ij} \ ,
\end{eqnarray}
\begin{eqnarray}\label{eq:2}
	-\mathcal{H}B^{(2)} -\frac{1}{2}B^{(2)'} -\frac{1}{2}\phi^{(2)}+\frac{1}{2}\psi^{(2)}=- \Delta^{-1}\left(\partial^{i} \Delta^{-1} \partial^{j}-\frac{1}{2} \mathcal{T}^{ij}\right) S^{(2)}_{ij} \ ,
\end{eqnarray}
\begin{eqnarray}\label{eq:3}
	\partial^i\left( \mathcal{H}\partial_i\phi^{(2)}+\partial_i\psi^{(2)'}+S^{(2)}_i \right)=0 \ ,
\end{eqnarray}
where $S^{(2)}_{ij}$ and $S^{(2)}_{i}$ are second order source terms
\begin{eqnarray}\label{eq:Sij}
	S^{(2)}_{ij}&&=\delta_{ij}\left( -8\mathcal{H}\phi\phi'-4\mathcal{H}\phi'\psi-12\mathcal{H}\phi\psi'-2\phi'\psi'-4\phi\psi''-\frac{16}{3}\mathcal{H}\phi\Delta B-\phi'\Delta B-\frac{5}{3}\psi'\Delta B\right. \nonumber\\
	&&\left.-2\phi\Delta B'-2\phi\Delta \phi-\frac{10}{3}\psi\Delta \psi+2\mathcal{H}\partial_bB'\partial^b B-2\mathcal{H}\partial_b\phi\partial^b B-\frac{2}{3}\mathcal{H}\partial_b\psi\partial^bB-2\partial_b\psi'\partial^b B\right.\nonumber\\
	&&\left.-\frac{2}{3}\partial_b \phi \partial^b\phi+\frac{2}{3\mathcal{H}}\partial_b \psi' \partial^b\phi-3\partial_b\psi \partial^b\psi+\frac{1}{3\mathcal{H}^2}\partial_b \psi' \partial^b\psi'-\frac{2}{3}\Delta B\Delta B+\frac{2}{3}\partial_c\partial_b B\partial^b\partial^cB \right) \nonumber\\
	&&-\partial_b\partial_jB\partial^b\partial_i B-2\mathcal{H}\partial_i\psi\partial_jB-\partial_i\psi'\partial_jB-\partial_i\psi \partial_j B'-\partial_i\psi \partial_j \phi -\frac{1}{\mathcal{H}}\partial_i\psi'\partial_j\phi-2\mathcal{H}\partial_iB\partial_j \psi \nonumber\\
	&&-\partial_iB'\partial_j \psi-\partial_i\phi\partial_j\psi+3\partial_i\psi\partial_j\psi-\partial_iB\partial_j\psi'-\frac{1}{\mathcal{H}}\partial_i\phi \partial_j\psi'-\frac{1}{\mathcal{H}^2}\partial_i\psi'\partial_j\psi'+4\mathcal{H}\phi\partial_i\partial_j B \nonumber\\
	&&+\phi'\partial_i\partial_j B+\psi'\partial_i\partial_j B+\Delta B\partial_i\partial_j B +2\phi\partial_i\partial_j B'+2\phi \partial_i\partial_j\phi+2\psi\partial_i\partial_j \psi \ ,
\end{eqnarray}
\begin{eqnarray}\label{eq:Si}
		S^{(2)}_{i}&&=\frac{1}{2\mathcal{H}^2}\partial_b\partial_iB'\partial^b B+\frac{1}{2\mathcal{H}^2}\partial_b\partial_i\phi\partial^b B-\frac{1}{2\mathcal{H}^2}\partial_b\partial_iB\partial^b \phi+2\phi\partial_iB-\frac{1}{\mathcal{H}}\phi'\partial_iB-4\psi\partial_iB \nonumber\\
		&&-\frac{3}{\mathcal{H}}\psi'\partial_i B-\frac{1}{\mathcal{H}^2}\psi''\partial_i B -\frac{4}{3\mathcal{H}}\Delta B\partial_i B - \frac{1}{2\mathcal{H}^2}\Delta B'\partial_i B-\frac{1}{2\mathcal{H}^2}\Delta \phi \partial_i B\nonumber\\
		&&+\frac{4}{3\mathcal{H}^2}\Delta \psi\partial_i B+\frac{1}{\mathcal{H}}\phi\partial_i \phi-\frac{2}{\mathcal{H}}\psi\partial_i\phi-\frac{1}{\mathcal{H}^2}\psi'\partial_i\phi-\frac{1}{6\mathcal{H}^2}\Delta B\partial_i \phi+\frac{2}{3\mathcal{H}^3}\Delta \psi\partial_i \phi \nonumber\\
		&&-\frac{2}{\mathcal{H}^2}\psi'\partial_i \psi -\frac{1}{\mathcal{H}^2}\phi\partial_i \psi' - \frac{4}{\mathcal{H}^2}\psi \partial_i \psi'-\frac{2}{\mathcal{H}^3}\psi'\partial_i \psi' -\frac{2}{3\mathcal{H}^3}\Delta B\partial_i \psi' \nonumber\\
		&&+\frac{2}{3\mathcal{H}^4}\Delta \psi\partial_i \psi'  \ .
\end{eqnarray}
 $\mathcal{T}^{ij}$ is defined as $\mathcal{T}^{ij}=\delta^{ij}-\partial^{i} \Delta^{-1} \partial^{j}$. The second order energy density perturbations $\delta^{(2)}=\rho^{(2)}/\rho^{(0)}$ can be calculated in terms of second order induced scalar perturbations and first order scalar perturbations 
\begin{eqnarray}\label{eq:4}	
	\delta^{(2)}&=&-2\phi^{(2)}+\frac{2}{3\mathcal{H}^2}\Delta \psi^{(2)}-\frac{2}{3\mathcal{H}}\Delta B^{(2)}-\frac{2}{\mathcal{H}}\psi^{(2)'}+S^{(2)}_{\rho} \ , 
\end{eqnarray}
where $S^{(2)}_{\rho}$ are given by
\begin{eqnarray}\label{eq:Sr}	
	S^{(2)}_{\rho}&&= 8\phi^2 +\frac{8}{\mathcal{H}} \phi\psi' -\frac{8}{\mathcal{H}} \psi\psi'+\frac{2}{\mathcal{H}^2} (\psi')^2+\frac{8}{3\mathcal{H}}\phi\Delta B -\frac{8}{3\mathcal{H}} \psi\Delta B+\frac{4}{3\mathcal{H}^2} \psi'\Delta B \nonumber\\
	&&+\frac{16}{3\mathcal{H}^2} \psi\Delta \psi-2\partial_b B\partial^b B+\frac{4}{3\mathcal{H}} \partial_b\psi\partial^b B-\frac{2}{3\mathcal{H}^2}  \partial_b\phi\partial^b\phi-\frac{4}{3\mathcal{H}^3} \partial_b\psi'\partial^b \phi \nonumber\\
	&& +\frac{2}{\mathcal{H}^2}\partial_b\psi\partial^b\psi -\frac{2}{3\mathcal{H}^4}\partial_b\psi'\partial^b \psi' + \frac{1}{3\mathcal{H}^2} \Delta B\Delta B-\frac{1}{3\mathcal{H}^2}\partial_c\partial_b B\partial^c\partial^b B \ .
\end{eqnarray}

\section{Upper bounds on primordial curvature perturbation}\label{sec:3.0}
Since the large-amplitude primordial curvature perturbations $\zeta_{\mathbf{k}}$ on small scales are necessary for the formation of primordial black holes, we consider the log-normal primordial power spectrum
\begin{eqnarray}\label{eq:P}
	\mathcal{P}_{\zeta}(k)=\frac{A_{\zeta}}{\sqrt{2 \pi \sigma_*^2}} \exp \left(-\frac{\ln \left(k / k_*\right)^2}{2 \sigma_*^2}\right) \ .
\end{eqnarray}
In the limit of $\sigma_* \rightarrow 0$, $\mathcal{P}_{\zeta}(k)$ approaches to a  monochromatic power spectrum, namely $\mathcal{P}_{\zeta}(k)=A_\zeta k_* \delta\left(k-k_*\right)$.

In previous studies on primordial black holes, the \ac{PDF} of the primordial curvature perturbation $\zeta_{\mathbf{k}}$ was often assumed to be Gaussian, resulting in a Gaussian \ac{PDF} for the first order energy density perturbation $\delta^{(1)}$ \cite{Sato-Polito:2019hws}. In this Letter, we consider the contribution of second order scalar induced energy density perturbations $\delta^{(2)}$ to the formation of primordial black holes. As we mentioned, the statistics of second order scalar induced energy density perturbations are highly non-Gaussian. Therefore, the \acp{PDF} of the total energy density perturbation $\delta_r=\delta^{(1)}+\frac{1}{2}\delta^{(2)}$ cannot be directly and simply obtained. We follow the method described in Ref.~\cite{Nakama:2016enz} to calculate the highly non-Gaussian \acp{PDF} of $\delta_r$. In Ref.~\cite{Nakama:2016enz}, the \acp{PDF} of the second order energy density perturbation induced by primordial gravitational waves was studied. Here, we calculate the \acp{PDF} of total energy density perturbation $\delta_r=\delta^{(1)}+\frac{1}{2}\delta^{(2)}$ induced by primordial curvature perturbations $\zeta_{\mathbf{k}}$. The relation between $\delta_r$ and $\zeta_{\mathbf{k}}$ can be rewritten as
\begin{equation}\label{eq:A}
	\begin{aligned}
		\delta_r &= \delta^{(1)}+\frac{1}{2}\delta^{(2)} \\
		&=  \frac{2(dk)^3}{3(2\pi)^{3/2}}\sum_{\mathbf{k}_i}\frac{\mathcal{P}_{\zeta}(k_*|\mathbf{k}_i|)}{A_{\zeta}}W(|\mathbf{k}_i|R)\zeta(\mathbf{k}_i)\delta^{(1)}(|\mathbf{k}_i|\eta)\\
		&+\frac{2(dk)^6}{9(2\pi)^3}\sum_{\mathbf{k}_i,\mathbf{k}_j}\frac{\mathcal{P}_{\zeta}(k_*|\mathbf{k}_i|)}{A_{\zeta}}\frac{\mathcal{P}_{\zeta}(k_*|\mathbf{k}_j|)}{A_{\zeta}}W(|\mathbf{k}_i+\mathbf{k}_j|R)\\
		&\times\zeta(\mathbf{k}_i)\zeta(\mathbf{k}_j)I_{\delta^{(2)}}(\mathbf{k}_i,\mathbf{k}_j,\eta)  \ ,
	\end{aligned}
\end{equation}
where $W(x)=3\left(\sin x-x\cos x \right)/x^3$ is the window function \cite{Nakama:2016enz}. $I_{\delta^{(2)}}$ is the kernel function of $\delta^{(2)}$  \footnote{We provid the specific expression for $I_{\delta^{(2)}}$ in the Python program available on the following website:https://github.com/MrFastThree/Idelta-Python-Calculation-Program.git.}. And $\mathcal{P}_{\zeta}$ is the log-normal primordial power spectrum. Eq.~(\ref{eq:A}) describes the relationship between the random variables $\delta_r$ and $\zeta$. The \ac{PDF} of $\delta_r$ (non-Gaussian distribution) can be simulated in terms of the PDF of $\zeta$. The mass-scale relation of PBHs can be expressed as follows \cite{Nakama:2016enz}:
\begin{equation}
	\begin{aligned}
  M_{\mathrm{PBH}} = 2.2 \times 10^{13} M_{\odot}\left(\frac{k}{1\mathrm{Mpc}^{-1}}\right)^{-2}\ .
	\end{aligned}
\end{equation}

\begin{figure}
	\centering
	\includegraphics[scale = 0.55]{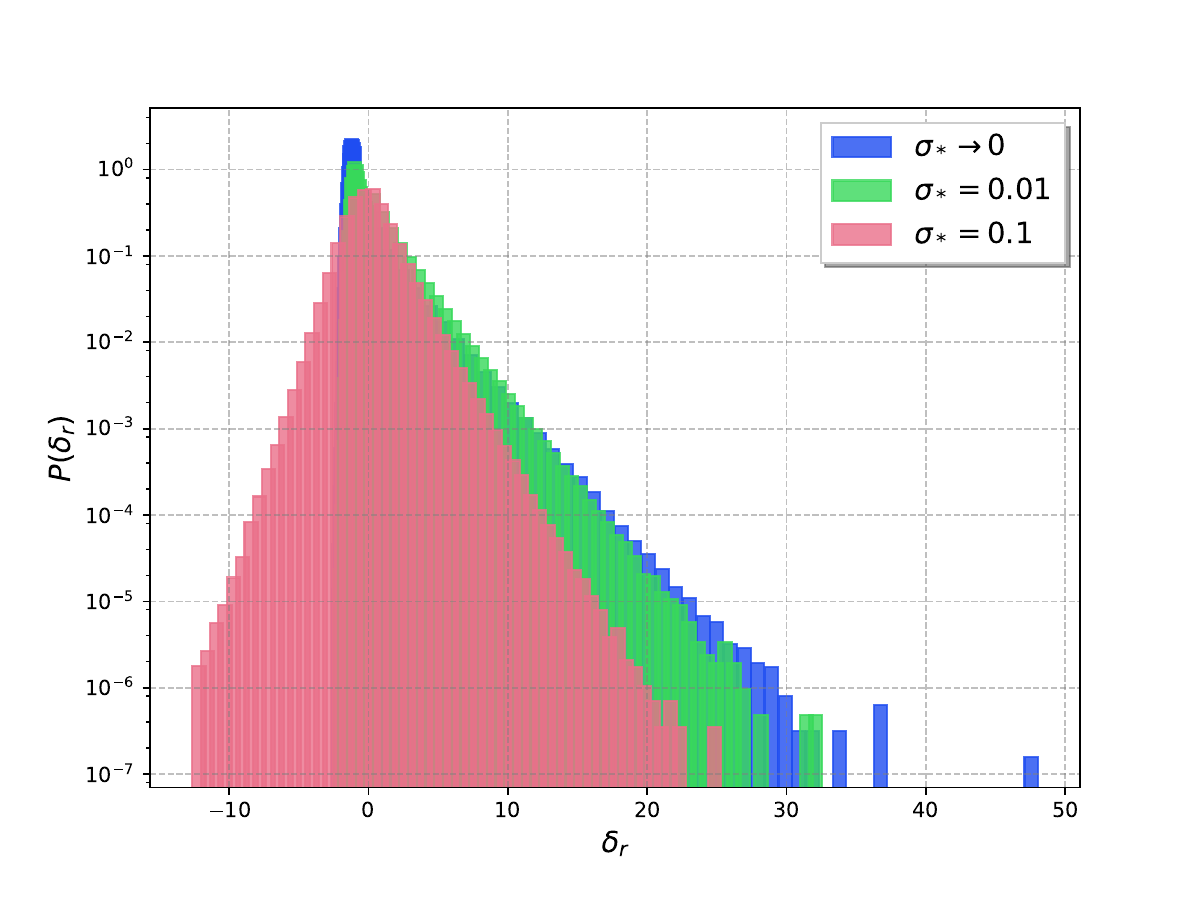}
	\caption{\acp{PDF} of total energy density perturbation $\delta_r=\delta^{(1)}+\frac{1}{2}\delta^{(2)}$ for log-normal primordial power spectrum with different $\sigma_*$, where we have set $A_\zeta = 1$ and $\eta\approx0.4$.}\label{fig:pdf}
\end{figure}

In order to intuitively show the trend of \acp{PDF} with respect to $\sigma_*$, we calculate \acp{PDF} corresponding to different $\sigma_*$ values near $\sigma_*=0$. As shown in Fig.~\ref{fig:pdf}, the \acp{PDF} of $\delta_r$ are highly non-Gaussian and contract gradually as $\sigma_*$ increases. Therefore, relying solely on the variance, $\sigma^2$, is insufficient to evaluate the probability of primordial black hole formation. In this paper, we calculate the PDF function corresponding to different values of $\eta$ within the range $[0.2, 10]$. Similar to Reference \cite{Nakama:2016enz}, we select the moment that yields the maximum number of \acp{PBH}, resulting in a final $\eta$ of approximately 0.4.
The relative abundance of \acp{PBH} $f_{\mathrm{pbh}}(m_{\mathrm{pbh}})$ is equivalent to the probability that the smoothed density field exceeds the 
threshold $\delta_c$ \cite{Press:1973iz,Sato-Polito:2019hws}, 
\begin{eqnarray}\label{eq:beta}
	\beta\left(M_{\mathrm{PBH}}\right)= \int_{\delta_c} d \delta_r P(\delta_r) \ ,
\end{eqnarray}
where the value of the threshold $\delta_c$ is, in fact, influenced by the characteristics of the primordial perturbation (its shape and the degree of non-Gaussianity). In reality, this value may vary significantly depending on these properties \cite{Musco:2018rwt, Musco:2020jjb,NobleChamings:2019ody,DeLuca:2019qsy,DeLuca:2023tun}. 
Here, we are primarily concerned with the impact of introducing second-order density perturbations on the \ac{PDF} of the final density perturbation. It is important to emphasize that our calculation of the PDF here does not depend on the choice of the threshold $\delta_c$. For simplicity, we have set $\delta_c = 0.4$.

\begin{figure}
	\centering
	\includegraphics[scale = 0.55]{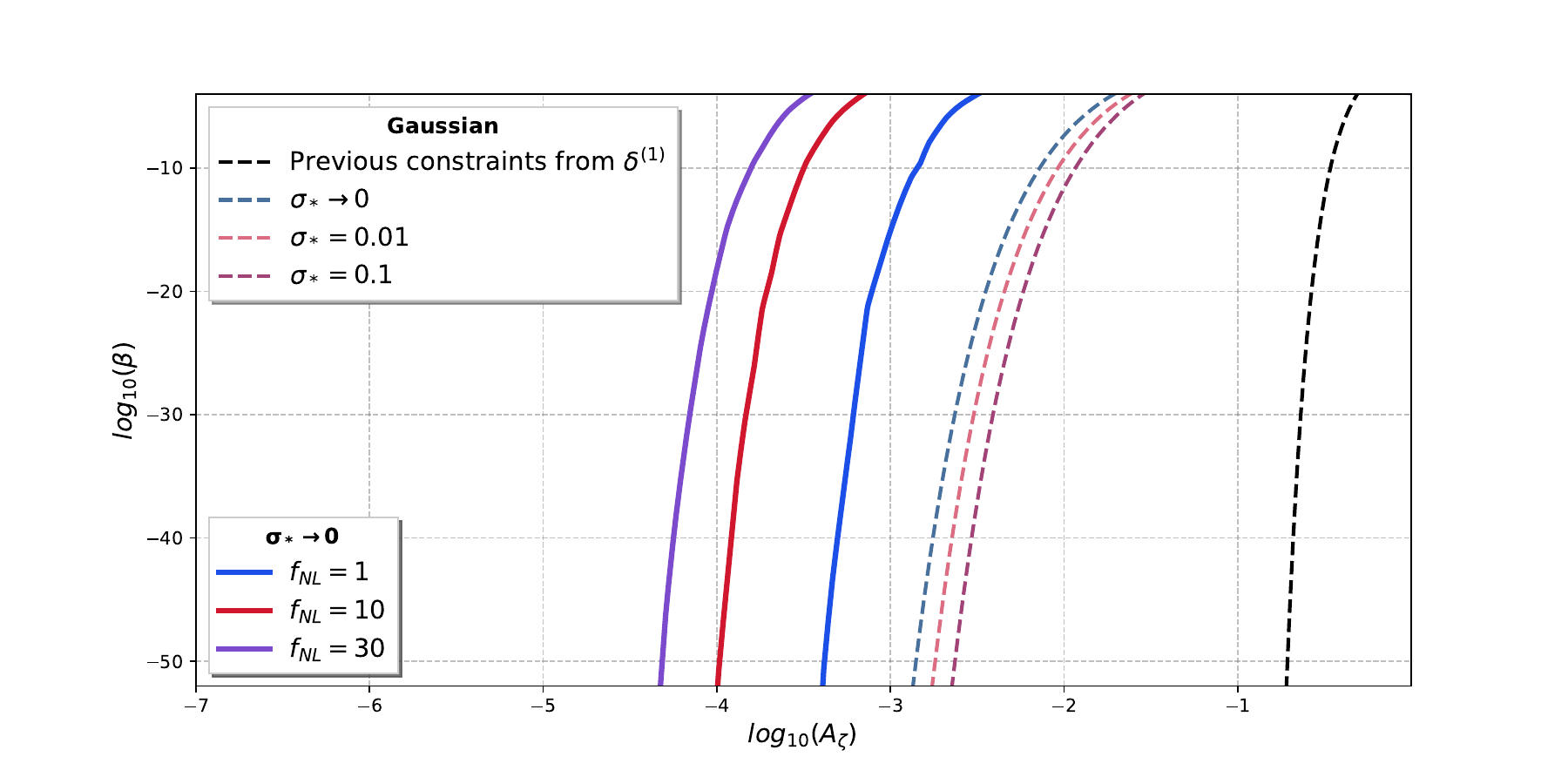}
	\caption{ The plot of the primordial black hole abundance $\beta$ as a function of the amplitude of the primordial power spectrum $A_\zeta$. Three colored dashed lines represent the relations between $\beta$ and $A_{\zeta}$ for different $\sigma_*$. Three solid lines represent $\beta(A_{\zeta})$ for the local-type non-Gaussian primordial curvature perturbations $\zeta^{NG}_{\mathbf{k}}$ with different $f_{\mathrm{NL}}$. The black dashed line represents the previous results from $\delta^{(1)}$ in Ref.~\cite{Sato-Polito:2019hws}.}\label{fig:beta}
\end{figure}

\begin{figure}[htbp]
	\subfloat[]{\includegraphics[scale = 0.38]{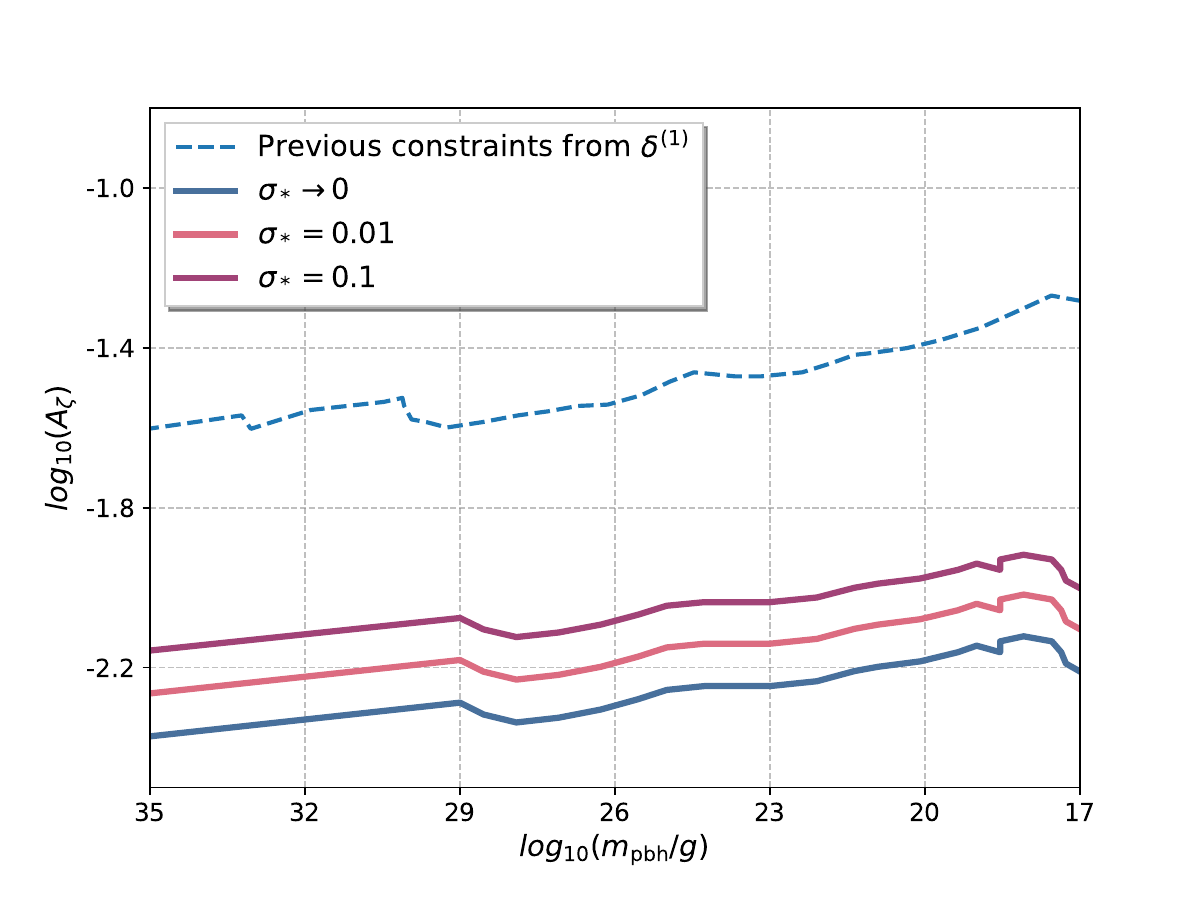}\label{fig:ng}}
	\subfloat[]{\includegraphics[scale = 0.38]{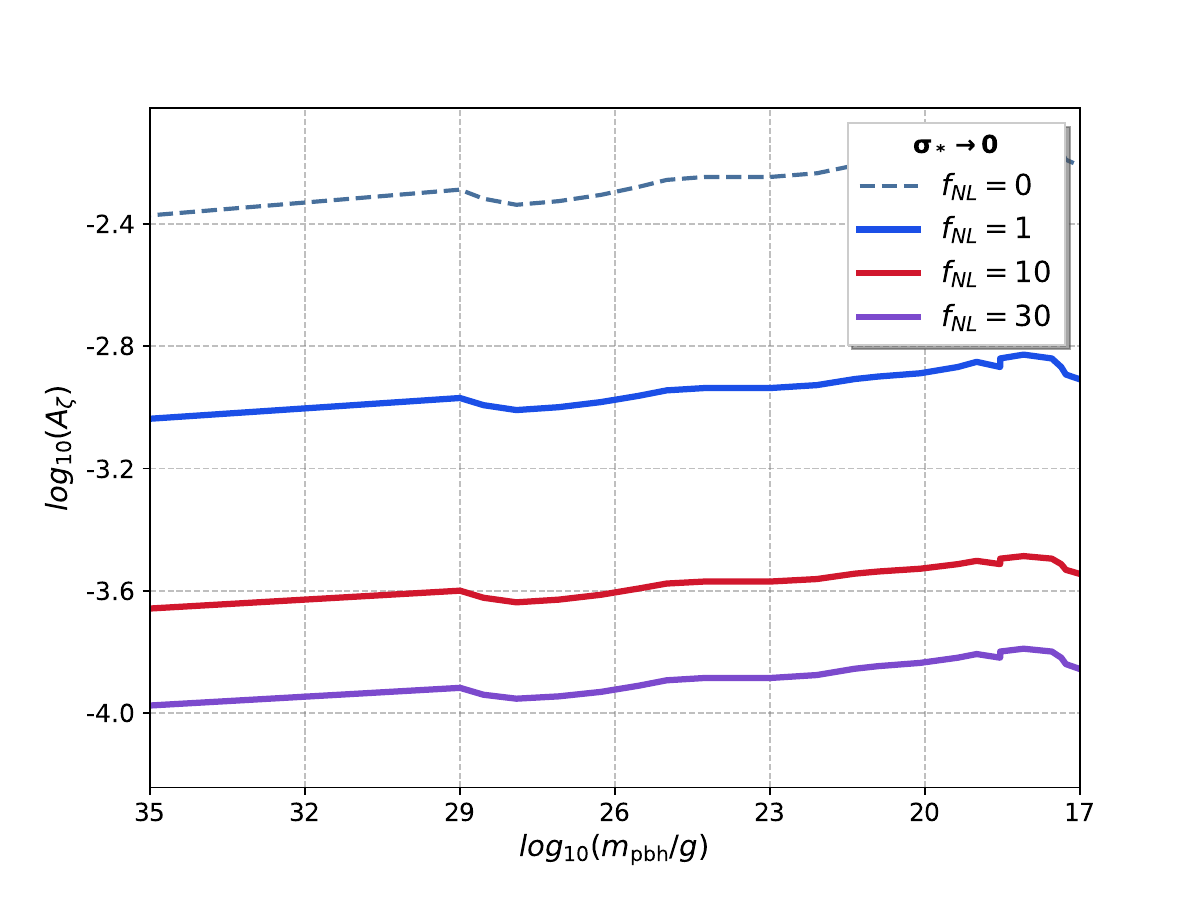}\label{fig:fnl}}
	\caption{\textbf{Left panel}: The upper bounds of $A_{\zeta}$ from total energy density perturbation $\delta_r$ with different $\sigma_*$. The blue  dashed line represents the previous results from $\delta^{(1)}$ in Ref.~\cite{Sato-Polito:2019hws}. \textbf{Right panel}: The upper bounds of $A_{\zeta}$ for different $f_{\mathrm{NL}}$. The dashed line represents the case of Gaussian primordial curvature perturbation.}
\end{figure}

By substituting \ac{PDF} $P\left(\delta^{(1)}+\frac{1}{2}\delta^{(2)}\right)$ of total energy density perturbations into Eq.~(\ref{eq:beta}), we obtain $\beta(A_{\zeta})$ as a function of $A_{\zeta}$ (Fig.~\ref{fig:beta}). Since $\beta$ has been constrained at different masses of \acp{PBH}, we use Fig.~18 in Ref.~\cite{Carr:2020gox} to determine the upper bounds of $A_{\zeta}$. As shown in Fig.~\ref{fig:ng}, for log-normal primordial power spectrum, its amplitude $A_{\zeta}$ is constrained to be about $A_{\zeta}\sim 3\times10^{-3}$.
The local-type non-Gaussian primordial curvature perturbation is given by \cite{Cai:2018dig,Li:2023xtl}
\begin{eqnarray}\label{eq:Png}
	\zeta^{NG}_{\mathbf{k}}=\zeta^G_{\mathbf{k}}+\frac{3}{5}f_{\mathrm{NL}}\int\frac{d^3n}{(2\pi)^{3/2}} \zeta^G_{\mathbf{k}-\mathbf{n}}\zeta^G_{\mathbf{n}}  \ ,
\end{eqnarray}
where $\zeta^G$ is the Gaussian primordial curvature perturbation. Here, we consider the lowest order contributions of local-type non-Gaussianity. Namely, $\delta^{(1)}_{NG}\sim A_{\zeta}f_{\mathrm{NL}}$ and $\delta^{(2)}_{NG}\sim A^{3/2}_{\zeta}f_{\mathrm{NL}}$. As shown in Fig.~\ref{fig:beta}, in this case, $\beta\left(A_{\zeta},f_{\mathrm{NL}} \right)$ is a function of $A_{\zeta}$ and $f_{\mathrm{NL}}$. The upper bounds of $A_{\zeta}$ with different $f_{\mathrm{NL}}$ are given in Fig.~\ref{fig:fnl}. For $f_{\mathrm{NL}}=10$, the upper bound of $A_{\zeta}$ is about $2.5\times10^{-4}$. When we consider the second order induced energy density perturbation, the effects of primordial non-Gaussianity will greatly reduce the upper bounds of $A_{\zeta}$ even for a relatively small $f_{\mathrm{NL}}$.

\section{Conclusion}\label{sec:4.0}
In this Letter, we have analyzed the new effects of second order scalar induced energy density perturbation $\delta^{(2)}$ on \acp{PBH} formation. We calculated \acp{PDF} and $\beta$ as a function of $A_{\zeta}$ and $\sigma_*$ in terms of the log-normal primordial power spectra, where we have chosen the threshold $\delta_c=0.4$ for convenience. In order not to overproduce \acp{PBH}, the parameters of primordial power spectra are constrained. For $\sigma_* \to 0$, the amplitude of primordial power spectra is constrained to be about $A_{\zeta}\sim 3\times10^{-3}$. With the increase of $\sigma_*$, the upper bounds of $A_{\zeta}$ will gradually increase. The local-type non-Gaussianity is also considered in this Letter. For $f_{\mathrm{NL}}=10$, the upper bound of $A_{\zeta}$ is about $2.5\times10^{-4}$. The effects of primordial non-Gaussianity will greatly reduce the upper bounds of $A_{\zeta}$ even for a relatively small $f_{\mathrm{NL}}$. In addition, since the upper bound of $A_{\zeta}$ is constrained to be about $A_{\zeta}\sim 3\times 10^{-3}$ for Gaussian primordial curvature perturbation, the energy density of third order \acp{SIGW} is always less than that of second order \acp{SIGW}.

The relative abundance of \acp{PBH} $f_{\mathrm{pbh}}(m_{\mathrm{pbh}})$ depends on the probability that the smoothed density field exceeds the 
threshold $\delta_c$ \cite{Press:1973iz}, namely, $\beta\left(M_{\mathrm{PBH}}\right)= \int_{\delta_c} d \delta_r P(\delta_r)$. To investigate the impact of second order induced density perturbation $\delta^{(2)}$ on $\beta$, we first need to calculate the corresponding \acp{PDF} and the threshold $\delta_c$. The calculation of the threshold $\delta_c$ is quite complex, as it depends on the shape of the primordial power spectrum and the specific primordial black hole model under consideration. In this article, we treat the threshold $\delta_c$ as a fixed parameter, focusing on the impact of second-order scalar-induced density perturbations on the \acp{PDF}. The effects of higher-order density perturbations on the threshold $\delta_c$ may be studied in future work.

\section*{Acknowledgement}
This work has been funded by the National Nature Science Foundation of China under grant No. 12475075, No. 12075249, and 11690022, and the Key Research Program of the Chinese Academy of Sciences under Grant No. XDPB15. We acknowledge the \texttt{xPand} package \cite{Pitrou:2013hga}. 

\bibliography{biblio}

\end{document}